\documentclass[twocolumn,
showpacs,preprintnumbers]{revtex4}
\usepackage{graphicx}

\begin{document}
\preprint{\vtop{{\hbox{YITP-03-75}\vskip-0pt
}}}
\thispagestyle{empty}
\title{On the charmed scalar resonances 
}
\author{K. Terasaki}
\affiliation{ Yukawa Institute for Theoretical Physics, 
Kyoto University, Kyoto 606-8502, Japan
}
\thispagestyle{empty}
\begin{abstract}
It is discussed that a newly observed broad bump around $2.3$ GeV in 
the $D\pi$ mass distribution can include two scalar resonances, and 
is predicted that a broad scalar resonance can be observed around 
$\sim 2.45$ GeV in the $DK$ channel. 
\end{abstract}

\vskip 0.5cm
\maketitle

Recently, it has been reported that a charmed scalar resonance 
$D_0^0(2.31)$ with a mass $2308\pm 60$ MeV and a width 
$279 \pm 99$ MeV has been observed~\cite{BELLE-D_0}, and has been 
suggested that it is an ordinary $^3P_0$ $\{c\bar u\}$ meson. 

If it is the case, it can be compared with the $K_0^*$ which has been
considered as the $^3P_0\,\,\{n\bar s\},\,\,(n=u,\,d),$ 
state~\cite{CT}. However, the resulting width is much narrower than
the observed one as seen  later. In addition, its strange counterpart
is expected to be  observed around $\sim 2.4$ GeV from a quark
counting with the mass difference $\Delta m_s = m_s - m_n\simeq 0.1$
GeV, ($n=u,\,d$), as a broad resonance since its mass is higher than 
the $DK$ threshold. However, such a resonance has not been confirmed. 
Therefore, another option might be to assign the $D_0(2.31)$ to the 
non-strange counterpart of the previously observed 
$D_{s0}^+(2.32)$~\cite{BABAR,CLEO,BELLE-D_s} which has been assigned
to various hadron
states~\cite{Terasaki-D_s,Terasaki-yitpws,Terasaki-hadron03,models}. 
However, the mass of the $D_{s0}^+(2.32)$ is very close to that of 
the non-strange $D_0(2.31)$ under consideration although the 
$D_{s0}^+(2.32)$ is expected to be heavier by $\Delta m_s \simeq 0.1$
GeV than the non-strange $D_0(2.31)$ since the former contains an $s$
quark. For example, it might be assigned to the lighter class of the
scalar four-quark $\hat D_0 \sim [cn][\bar u\bar d]$ meson as the 
$D_{s0}^+(2.32)$ has been assigned to the $I_3=0$ component, 
$\hat F_I^+$, of the iso-triplet $\hat F_I\sim [cn][\bar s\bar n]$ 
mesons~\cite{Terasaki-D_s,Terasaki-yitpws,Terasaki-hadron03}, where 
the square brackets implies anti-symmetry with respect to the 
exchange of the flavors between them. In this case, however, its 
width should be much 
narrower~\cite{Terasaki-yitpws,Terasaki-hadron03} than the measured
width of the $D_0(2.31)$ because of the small overlapping of the 
wave functions between the parent scalar four-quark meson and the 
final two pseudoscalar meson states with respect to the color and 
spin~\cite{Jaffe}. In this way, it is seen that the properties of the 
$D_0(2.31)$ and the $D_{s0}^+(2.32)$ are very much different from 
each other, and is suspected that the quark content of the $D_0(2.31)$ 
is much different from the one of $D_{s0}^+(2.32)$ and there 
coexist two resonances with different structure in the region 
of the broad bump. In this short note, therefore, we will 
demonstrate that, in the region of the very broad $D_0(2.31)$ bump, 
two resonance peaks can be included, i.e., one is the 
ordinary scalar $D_0^*\sim \{c\bar n\}$  meson and the other is the 
scalar four-quark $\hat D \sim [cn][\bar u\bar d]$ meson with a 
narrow width, although the BELLE collaboration have assumed an 
amplitude  which includes only one scalar resonance in addition to 
the other resonances with different spins~\cite{BELLE-D_0}. Besides,  
we study the strange counterpart $D_{s0}^{*+}$ of the $D_0^*$. 

We start with the general form of decay rate,  
\begin{eqnarray}
\Gamma(A \rightarrow B + \pi)&&
=\Biggl({1\over 2J_A + 1}\Biggr)  
\Biggl({q_c\over 8\pi m_A^2}\Biggr)  \nonumber\\
&&\times
\sum_{spins}|M(A \rightarrow B + \pi)|^2 , 
\label{eq:rate}
\end{eqnarray}
for a two body decay, 
$A({\bf p}) \rightarrow B({\bf p'}) + \pi({\bf q})$,
where $J_A$, $q_c$ and $M(A \rightarrow B + \pi)$ denote the spin 
of the parent $A$, the center-of-mass momentum of the final $B$ and 
$\pi$ mesons, and the decay amplitude, respectively. To calculate 
the amplitude, we use the PCAC (partially conserved axial-vector 
current) hypothesis and a hard pion approximation in the infinite 
momentum frame (IMF), i.e., ${\bf p}\rightarrow\infty$~\cite{suppl}. 
In this approximation, the amplitude is evaluated at a little
unphysical point, i.e., $m_\pi^2 \rightarrow 0$, and is approximately
given by
\begin{equation}
M(A \rightarrow B + \pi) 
\simeq \Biggl({m_A^2 - m_B^2\over f_\pi}\Biggr)
\langle{B|A_{\bar \pi}|A}\rangle ,                         
\label{eq:amp}
\end{equation}
where $A_\pi$ is the axial counterpart of the isospin, $I(=V_\pi)$. 
The {\it asymptotic matrix element} of $A_\pi$ (matrix elements of 
$A_\pi$ taken between single hadron states with infinite momentum), 
$\langle{B|A_\pi|A}\rangle$, gives the dimensionless $AB\pi$ 
coupling strength. 

We now consider the $K_0^{*}\rightarrow K\pi$ decays and estimate the 
size of the asymptotic matrix element of $A_\pi$, where the $K_0^{*}$ 
has been considered as the $^3P_0\,\,\{n\bar s\},\,\,(n=u,d)$, state. 
Substituting the measured values~\cite{PDG02}, 
$\Gamma(K_0^{*}\rightarrow all)=294\pm 23$ MeV and 
${\rm Br}(K_0^{*}\rightarrow K\pi)=93 \pm 10\,\,\%$,  
into Eq.(\ref{eq:rate}) and using Eq.(\ref{eq:amp}), we obtain 
\begin{equation}
|\langle{K^+|A_{\pi^+}|K_0^{*0}}\rangle| \simeq 0.29, 
                                          \label{eq:AME-size}
\end{equation}
where we have used the iso-spin $SU_I(2)$ symmetry which is always 
assumed in this short note. 

Next, we consider the newly observed $D_0(2.31)\rightarrow D\pi$ 
decay. Insertion of the measured mass $\simeq 2.31$ GeV and width 
$\simeq 280$ MeV~\cite{BELLE-D_0} into Eq.(\ref{eq:rate}) and use of 
Eq.(\ref{eq:amp}) with the $SU_I(2)$ symmetry  
lead to 
$|\langle{D^+|A_{\pi^+}|D_0^{0}(2.31)}\rangle| \simeq 0.58$.  
It is too big. It is larger by about factor two than the  
$|\langle{K^+|A_{\pi^+}|K_0^{*0}}\rangle|$, although 
({\it asymptotic}) $SU_f(4)$ symmetry (roughly speaking, $SU_f(4)$ 
symmetry of asymptotic matrix elements) would 
predict~\cite{suppl,Hallock}  
\begin{equation}
\langle{D^+|A_{\pi^+}|D_0^{0}(2.31)}\rangle 
=\langle{K^+|A_{\pi^+}|K_0^{*0}}\rangle,   \label{eq:AME-Api-sym}
\end{equation}
if the $D_0^{}(2.31)$ were a pure two-quark system, 
$D_0^{}(2.31)\sim \{c\bar n\}$, $(n=u,\,d)$, as the 
$K_0^*\sim \{n\bar s\}$ mentioned before. 
(Asymptotic flavor symmetry and 
its fruitful results were reviewed in Ref.~\cite{suppl}.) The 
measure of the (asymptotic) flavor symmetry breaking is given by 
the form factor $f_+(0)$ of related vector current. The estimated 
values of $f_+(0)$'s of charm changing currents have been compiled 
in Ref.~\cite{PDG94} as 
\begin{eqnarray}
f_+^{(\bar K D)}(0)&&\simeq 0.75 \pm 0.02 \pm 0.02, \nonumber\\
{f_+^{(\pi D)}(0)\over f_+^{(\bar K D)}(0)}
&&\simeq 1.0 ^{+0.3}_{-0.2} \pm 0.4, \hspace{11mm} 
{\rm [Mark\,\, III]},                              \nonumber\\
&&\simeq 1.3 \pm 0.2 \pm 0.1, \hspace{8mm} {\rm [CLEO].}  
\label{eq:FF}
\end{eqnarray}
The above values suggest that the (asymptotic) flavor $SU_f(3)$ 
symmetry works considerably well while the $SU_f(4)$ is broken 
to the extent of 20 -- 30 $\%$. Nevertheless, the equality of 
Eq.(\ref{eq:AME-Api-sym}) would still work if the $D_0^{}(2.31)$ 
were a simple two-quark system (as the $K_0^*$), since 
it has been obtained by inserting the commutation relation, 
$[V_{D_s^+}, A_{\pi^+}]=0$, 
between $\langle{D^+}|$ and $|{K_0^{*-}}\rangle$ in the IMF and by 
using the $SU_I(2)$ symmetry and the charge conjugation invariance.  

The mass of the $^3P_0\,\,\{c\bar s\}$ meson has been predicted 
in the region, $\sim 2.4$ -- $2.5$ GeV, from various approaches, for 
example, the potential model~\cite{potential}, the lattice QCD with 
a static heavy quark and $N_f=2$ sea quarks~\cite{Bali},  the 
relativistic lattice QCD with the $N_f=2$ sea quarks~\cite{UKQCD}
except for the light-cone oscillator model which predicts $2.328$
GeV~\cite{Zhou}. The mass of the scalar $\{c\bar n\}$ meson also has 
been expected to be in the region, $\sim 2.3$ -- 2.4 GeV, from the 
same approaches except for the Ref.~\cite{Zhou}. (However, it is
strange that the mass of the scalar $\{c\bar n\}$ meson has not been
studied in Ref.~\cite{Zhou}.) Therefore, the mass of the $D_0(2.31)$ 
is compatible with the predicted one of the scalar 
$D_0^* \sim \{c\bar n\}$ mesons. 
However, the decay width should be much narrower than the measured 
one as seen below. To estimate roughly the width of the $D_0^*$, 
we insert the asymptotic $SU_f(4)$ symmetry relation, i.e., the one 
that the $D_0^{0}(2.31)$ is replaced by the $D_0^{*0}$ in 
Eq.(\ref{eq:AME-Api-sym}), and the mass ($m_{D_0^*}\simeq 2.35$ GeV 
which is around the average of the predicted values mentioned before) 
of $D_0^*$ into the general form of the decay rate, 
Eq.(\ref{eq:rate}), with the approximate amplitude, Eq.(\ref{eq:amp}), 
and use Eq.(\ref{eq:AME-size}) as the input data. Then the estimated 
width is approximately $\Gamma_{D_0^*}\simeq 90$ MeV, since it 
is expected that the rate for the $D_0^*\rightarrow D\pi$ decays 
saturate approximately the total decay rate of $D_0^*$. 

On the other hand, we have studied the lighter class of scalar 
four-quark mesons, $[cq][\bar q\bar q],\,\,(q=u,\,d,\,s)$, in 
Refs.~\cite{Terasaki-D_s,Terasaki-yitpws,Terasaki-hadron03}. We have
assigned the $D_{s0}^+(2.32)$ to the $\hat F_I^+$ as mentioned 
before and have predicted an iso-doublet, $\hat D$, with a mass 
around $\simeq2.22$ GeV and a width $\sim 15$ MeV. (However, the 
numerical value of the width should not be taken too literally, 
i.e., a width between several and 15 MeV might be acceptable since 
the width of the $\hat F_I^+$ as the input data is still not 
definite~\cite{Terasaki-hadron03}.) Therefore, the $\hat D$ can 
make a narrow peak around $\sim 2.2$ GeV in the $D\pi$ mass
distribution, so that two peaks, i.e., the relatively broader 
$D_0^*\sim \{c\bar n\}$ and the narrower 
$\hat D\sim [cn][\bar u\bar d]$ can be included in the very broad  
bump of the $D_0(2.31)$ observed by the BELLE collaboration. 

Now we consider the strange counterpart, 
$D_{s0}^{*+}\sim \{c\bar s\}$, of the $D_0^*\sim \{c\bar n\}$. 
Taking the mass difference $\Delta m_s=m_s - m_n\simeq 0.1$ GeV, 
we estimate $m_{D_{s0}^{*}}\simeq 2.45$ GeV which is 
compatible with the predicted values from various approaches 
mentioned before. The asymptotic $SU_f(4)$ symmetry can relate 
asymptotic matrix elements,  
$\langle{D}|A_{\bar K}|{D_{s0}^{*+}}\rangle$'s, to the 
$\langle{K^+|A_{\pi^+}|K_0^{*0}}\rangle$~\cite{suppl,Hallock}, 
\begin{eqnarray}
&&
\langle{D^0|A_{K^-}|D_{s0}^{*+}(2.45)}\rangle 
=\langle{D^+|A_{\bar K^0}|D_{s0}^{*+}(2.45)}\rangle \nonumber\\
&&=\langle{K^+|A_{\pi^+}|K_0^{*0}}\rangle.   \label{eq:AME-Ak-sym}
\end{eqnarray}
Using Eqs.(\ref{eq:rate}), (\ref{eq:amp}), (\ref{eq:AME-size}) 
and (\ref{eq:AME-Ak-sym}), we obtain 
\begin{equation}
\Gamma(D_{s0}^{*+}(2.45)\rightarrow (DK)^+)
\simeq   70\,\,{\rm MeV},   
                                       \label{eq:width-D_{s0}^*}
\end{equation}
which dominates the width of the $D_{s0}^{*+}(2.45)$. The iso-spin 
non-conserving $D_{s0}^{*+}(2.45)\rightarrow D_s^+\pi^0$ decay will 
give negligibly small contribution. It is compatible with the fact 
that no scalar resonance has been observed in the region up to 
$\simeq 2.7$ GeV above the $D_{s0}^+(2.32)$ resonance in the 
$D_s^+\pi^0$ mass distribution~\cite{BABAR}, i.e., in our case, the 
narrow peak at $2.32$ GeV in the $D_s^+\pi^0$ mass distribution is 
dominated by the iso-spin conserving 
$\hat F_I^+\rightarrow D_s^+\pi^0$ 
decay (since the iso-spin non-conserving 
$"D_{s0}^+"\rightarrow D_s^+\pi^0$ 
decay is suppressed, where the $"D_{s0}^+"$ denotes any iso-singlet 
scalar meson with charm $C=1$ and strangeness $S=1$ such as 
$D_{s0}^{*+}$, $\hat F_0^+$, etc.). It should be noted that the CLEO 
collaboration~\cite{CLEO-Kubota} have observed a peak around $2.39$  
GeV in the $DK$ mass distribution but it has been taken away as a 
false peak arising from the decay, 
$D_{s1}(2536)\rightarrow D^*K\rightarrow D[\pi^0]K$,  
where the $\pi^0$ has been missed. However, we hope that it can 
involve the true resonance corresponding to the $D_{s0}^{*+}(2.45)$ 
or that the resonance can be observed by experiments with higher 
luminosities. 

So far, we have studied the charmed scalar four-quark 
$[cq][\bar q\bar q]$ mesons involving the $\hat F_I$ and $\hat D$ in 
addition to the ordinary $^3P_0\,\{c\bar q\}$ mesons, $D_0^*$ and
$D_{s0}^{*+}$. However, the $[cq][\bar q\bar q]$ mesons are only a
part of possible four-quark mesons which can be classified into the 
four types, 
$\{qq\bar q\bar q\} = [qq][\bar q\bar q] \oplus (qq)(\bar q\bar q) 
\oplus \{[qq](\bar q\bar q)\pm (qq)[\bar q\bar q]\}$, 
($q=u,\,d,\,s,\,c$), 
where the last two on the right-hand-side can have the spin-parity,  
$J^P=1^+$, at their lowest level while the first two can have 
$J^P=0^+,\,1^+,\,2^+$. Each of them can be again classified into two
classes, since there exist two ways to get color singlet 
$\{qq\bar q\bar q\}$ states, i.e., to take the $SU_c(3)$ 
${\bf \bar 3\times 3}$ or the ${\bf 6\times\bar 6}$ configuration. 
However, these two can mix with each other, so that they are 
classified into the heavier and lighter classes~\cite{Jaffe}. 
The former is dominated by the ${\bf 6\times\bar 6}$ of $SU_c(3)$ 
since the force between two quarks is repulsive when they form the 
{\bf 6}-plet of $SU_c(3)$ although the four-quark system still can 
be bound by the attractive force between a color singlet quark and 
anti-quark pair which is stronger by eight times than the above 
repulsive force. The latter is dominated by the color 
${\bf \bar 3\times 3}$ configuration since the force between the 
color ${\bf\bar 3}$-plet quark pair is attractive and stronger by 
four times than the repulsive force between the color octet quark 
and anti-quark pair~\cite{Hori}. 
Neglecting the mixing, we obtain the lighter class of 
mesons which involve the scalar $[cq][\bar q\bar q]$  discussed 
before, the $(cq)(\bar q\bar q)$ with $J^P=0^+,\,1^+,\,2^+$  and the 
$\{[cq](\bar q\bar q)\pm (cq)[\bar q\bar q]\}$ with $J^P=1^+$.  
Some of them are expected to be narrow as the $[cq][\bar q\bar q]$ 
discussed before but the other can have broader  widths. Therefore, 
the four-quark mesons can have a rich spectrum and  are very much 
interesting, so that the classification of the remaining  members of 
the lighter class of charmed four-quark mesons will be one of the 
interesting subjects in near future. 

Besides, the heavier class of the four-quark mesons are expected to 
be much heavier than the lighter ones, in particular, some of the 
heavier class of scalar four-quark mesons which consist of the light 
quarks can have masses close to the ones of the ordinary charm 
mesons, $D$ and $D_s^+$, as predicted in Ref.~\cite{Jaffe}, so that 
they can play an important role in hadronic weak decays of the $D$ 
and $D_s^+$, i.e., they can give large contributions to the decays 
through their pole amplitudes. In this way, we can obtain a possible 
solution to the long standing puzzle in hadronic weak 
decays~\cite{PDG02}, 
\begin{eqnarray}
&&{\Gamma(D^0\rightarrow K^+K^-)
                     \over \Gamma(D^0\rightarrow \pi^+\pi^-)}
\simeq 3,
\end{eqnarray}
consistently with the other two-body decays of charm 
mesons~\cite{charm88,charm93,charm99}. The lighter class of scalar 
four-quark mesons, in particular, $E_{\pi\pi}$ and $E_{\pi K}$ 
mesons, which have exotic quantum numbers, also can play an important 
role in the understanding of the $|\Delta {\bf I}|= 1/2$ rule  
violation in $K\rightarrow \pi\pi$ decays in consistency with the 
other weakly interacting processes such as the $K_L$-$K_S$ mass 
difference, the $K_L\rightarrow \gamma\gamma$ and the Dalitz decays 
of $K_L$~\cite{Terasaki01}. Therefore, hadronic weak interactions are
intimately related to hadron spectroscopy. In this sense, the
four-quark mesons are again interesting. 

In summary we have studied the new broad resonance $D_0(2.31)$ 
observed by the BELLE collaboration, and pointed out that the broad 
bump can have a structure including two peaks arising from the 
$\hat D(2.22)\sim [cn][\bar u\bar d]$ and 
$D_0^*(2.35)\sim\{c\bar n\}$, 
although the BELLE collaboration have analyzed their measured data 
by using the amplitude which involves only one scalar meson 
resonance in addition to various resonances with different spins. 
The $D_0^*(2.35)$ has been predicted to be rather broad 
($\sim 90$ MeV) by comparing with the observed rate for the 
$K_0^*\rightarrow K\pi$ and by using  the asymptotic $SU_f(4)$ 
symmetry while the $\hat D(2.22)$ has been previously predicted to 
be much narrower ($\sim {\rm several}$ -- $15$ MeV). 

The strange counterpart $D_{s0}^*\sim \{c\bar s\}$ of the above
$D_0^*(2.35)$ has been predicted to have a mass around $\sim 2.45$ 
GeV which is compatible with the existing predictions from various 
approaches. Its width has been approximately saturated by the 
$D_{s0}^{*+}\rightarrow (DK)^+$ decays and has been predicted to be 
$\sim 70$ MeV. 
 
Therefore, it is desired that the measured $D\pi$ mass distribution 
around the $D_0(2.31)$ is re-analyzed by using an amplitude 
including two scalar resonances. It is also awaited that the 
$D_{s0}^{*+}$ is observed in the $DK$ channels by experiments with 
high luminocities.

\section*{Acknowledgments}
The author would like to thank Professor D.~Cassel, Cornell 
University, for conversations through which this work was motivated. 
He also would like to appreciate Professor T.~Kugo, Professor 
K.~Shizuya, Professor R.~Sasaki, Professor T.~Onogi, Professor 
S.~Sugimoto and the other members of High Energy Physics Group, 
Yukawa Institute for Theoretical Physics, Kyoto University for 
discussions.  This work is supported in part by the Grant-in-Aid for 
Science Research, Ministry of Education, Science and Culture, Japan 
(No. 13135101). 


\end{document}